# Quantization of diamagnetic current in a superconducting ring with the Josephson point contact

S. I. Bondarenko, V. P. Koverya, A. V. Krevsun, and N. M. Levchenko

*B. Verkin Institute for Low Temperature Physics and Engineering of the National Academy of Sciences of Ukraine, Prospect Lenina 47, Kharkov 61103, Ukraine*

It was established experimentally that a critical value of the diamagnetic current, excited by an external magnetic field in a superconducting ring (with an inductance of about $\sim 10^{-6}$ H) with a Nb-Nb clamping point contact having the Josephson contact properties is a strictly periodic function of field strength, despite the complex microstructure of the clamping contact. The reasons of the periodic dependence are discussed on a basis of the interference model of diamagnetic current and quantized values of the circulating current in the microinterferometer formed by the clamping contact.

## INTRODUCTION

A diamagnetic current $I_L$ is excited in a superconducting ring subjected to a constant magnetic field $H_\perp$, perpendicular to its plane. If the ring has a "weak" part with a critical current value $I_c$ which is low compared to the rest of the ring, then the superconducting diamagnetic current cannot be higher than $I_c$. Of particular interest are properties of rings with a weak link with a large inductance of the ring (i.e. when the inductance is much larger than that of superconducting quantum interferometers), since they are little addressed in the literature. As a "weak" part it is preferable to use one of the types of contacts with properties of the Josephson contact,[1] because a theory of some of them is quite well developed.[2] In experimental investigations a technological aspect of the possibility to use one or another contact for a specific type of research is also important. In recent years, considerable attention in this area is being paid to manufacturing quality control of contacts,[3–6] as well as possibility of using them not only at liquid helium temperature, but at higher temperatures too.[7]

Previously, a dependence $I_L(H_\perp)$ was studied mainly in rings of quantum interferometers[2] with one or two Josephson contacts of a given type. It was found that if the condition $\Phi_0^2/2L_0 > kT$ ($\Phi_0$ is a quantum of magnetic flux through the interferometer ring, $L_0$ is the inductance of the interferometer ring, $k$ is the Boltzmann constant, $T$ is the ring temperature) is fulfilled, the diamagnetic current is quantized with a period $\Delta I_L = \Phi_0/L_0$. But if $\Phi_0^2/2L_0 < kT$, there is no quantization of the current. The latter condition corresponds to high values of the ring inductance ($L_0 > 10^{-9}$ H at $T = 4.2$ K).

Recently,[8] in our studies at $T = 4.2$ K upon passing a constant transport current through superconducting rings with inductance $L$ from $10^{-6}$ to $6 \cdot 10^{-6}$ H, which contain a clamping point contact with the properties of the Josephson contact, the quantization of the transport current in branches of such a ring has been found. This effect can be explained by a complex microstructure of the clamping contact, which is a tiny quantum interferometer with several microcontacts of the type of S–c–S connected in parallel. The mentioned contact is technologically simple to manufacture, and stable enough over time. The purpose of this paper is to elucidate reaction of a superconducting structure in the form of such a ring (without a transport current through it) on a magnetic field $H_\perp$, the more so as to our knowledge there are no works in this area.

## EXPERIMENTAL PROCEDURE

The ring with a clamping contact is sketched in Fig. 1. It is similar to that studied by us in Ref. 8, with the only difference that in the study of effect of a magnetic field on the ring a constant transport current $I$ was not passed through the point contact. To create a magnetic field a coil with a current that generates the field $H_\perp$ (not shown in Fig. 1) was placed above the ring surface. The ring was formed from a niobium (Nb) microwire with the diameter $d = 0.07$, and a contact was made by mechanical pressing of microwire ends to each other. The diameter of the ring was a few centimeters.

Some microwire of the ring was wound in the form of a coil on a fluxgate sensor for contactless measurement of the current in the ring from its magnetic field. The inductance $L$ of the ring together with the coil was from $10^{-6}$ to $6 \cdot 10^{-6}$ H for different specimens of the ring. During the experiment, the ring was located in liquid helium ($T = 4.2$ K), and a dependence of the current ($I_L$) in the ring on the magnetic field $H_\perp$, created by the current of the coil, $I_{H\perp}$, was measured. To reduce the effect of random external magnetic fields the area with the structure under investigation was screened by a ferromagnetic shield.

## EXPERIMENTAL RESULTS AND THEIR DISCUSSION

Fig. 2(a) shows a typical dependence $I_L(I_{H\perp})$ obtained from one specimen of the ring. It is seen that on increasing the current $I_{H\perp}$ in the coil, creating a magnetic field, from zero to 60 mA the diamagnetic current $I_L$ in the ring increases up to 20 mA. On further increasing $I_{H\perp}$ a periodic decrease of $I_L$ down to 14 mA and an increase again up to 20 mA appear. The direction of variation of the current $I_L$ in this case is shown in Fig. 2(a) by single arrows. Thus, a modulation depth of the current within the ring, $\Delta I_L$, is 6 mA. And the decrease of the current is stepwise. On the dependence $I_L(H_\perp)$ the current $I_{H\perp}$ is limited by the value of 180 mA but the dependence remains to be also periodic at large values of $I_{H\perp}$. In Fig. 2(a) the other peculiarity of variation of the current in the ring is also seen. It is expressed in the fact that on decreasing the current $I_{H\perp}$ from the value, corresponding to any jumps of the current $I_L$, down to zero, the current $I_L$ changes following one of trajectories shown by thin lines with double arrows. In particular, when the current $I_{H\perp}$ is turned off after the first jump, in the ring the current, equal to the value of the jump, $\Delta I_L = 6$ mA, but with opposite sign, freezes which corresponds to appearance of a paramagnetic current in the ring, i.e. the current which creates a magnetic field in the direction of the vector of the turned-off field, $H_\perp$. Reducing the current $I_{H\perp}$ to zero after the second jump in the diamagnetic current $I_L$ results in doubling the frozen paramagnetic current (and hence the frozen field in the ring too).

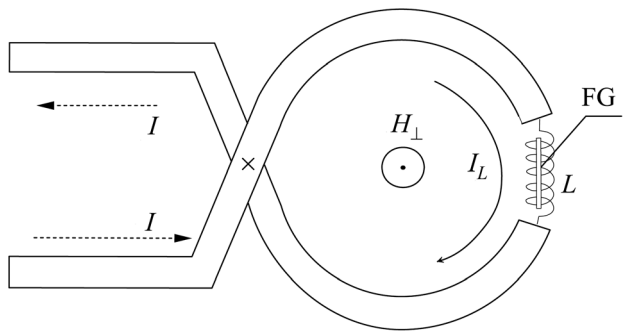

FIG. 1. A diagram of a superconducting ring with one point contact in a magnetic field $H_\perp$; FG is a sensor of flux gate; × is a position of the point contact between niobium wires, $I$ is the transport current in the case it passes through the point contact.

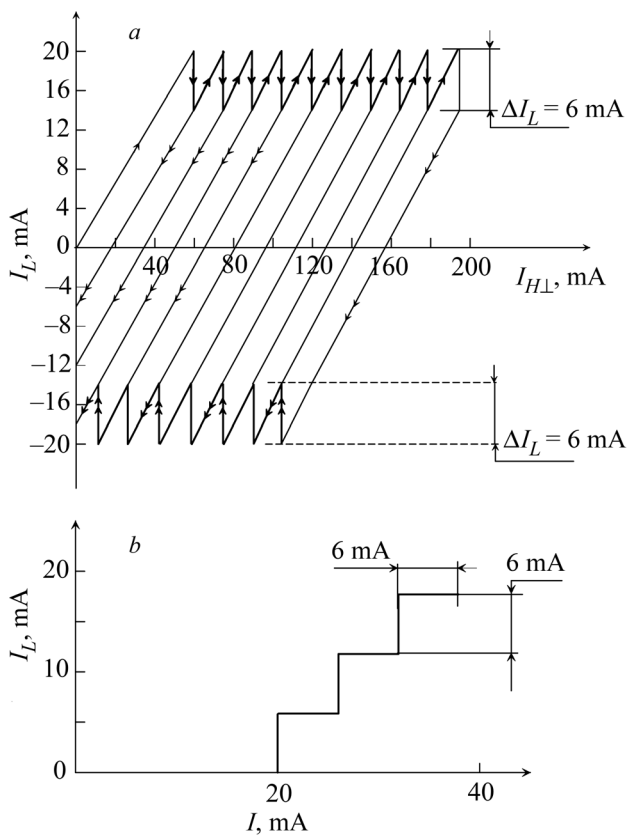

FIG. 2. A periodic dependence of the critical diamagnetic current (positive periodic values of the current $I_L$) and the critical paramagnetic current (negative periodic values of the current $I_L$) in a ring with a clamping point contact as a function of an external magnetic field $H_\perp$, perpendicular to the ring plane (a). A dependence of the current $I_L$ in the same ring on a magnitude of the transport current $I$, introduced to a point contact at $H_\perp = 0$ (b).

Reducing the current $I_{H\perp}$ after the other jumps almost does not increase the frozen current in the ring, because its value cannot be greater than the critical current of the contact. At the same time as the current $I_{H\perp}$ decreases after the third jump there appears a periodic dependence of the frozen current on a decreasing value of the current $I_{H\perp}$. The periodic dependence $I_L(I_{H\perp})$ in a negative range of values of $I_L$ is similar to the dependence $I_L(I_{H\perp})$ for positive values of $I_L$.

The dependence $I_L(I_{H\perp})$ is noteworthy due, first of all, to the strictly periodic variation of the critical diamagnetic and paramagnetic currents, and, second, to the value of the modulation depth of these currents. Regarding the periodicity, it should be noted that if to break the ring outside of a point contact and to study a magnetic-field dependence of the critical current of the isolated contact using a resistive technique, it turns out to be a complex amplitude-modulated and nonperiodic curve.[9] The reason for this dependence is a complex microstructure of the clamping contact which is found in a resistive operating mode of the contact. In our case, a critical state of the ring is formed by a constant magnetic field and is nondissipative. The dependence of the critical diamagnetic field on a magnetic field does not exhibit the abovementioned complexity and is strictly periodic. As for the modulation depth of critical currents, $\Delta I_L$, it appears to be equal to a height of periodic current steps in the dependence $I_L(I)$ for the same ring as shown in Fig. 2(b). A comparison of the dependencies $I_L(I_{H\perp})$ and $I_L(I)$ also shows that periodic jumps of the critical diamagnetic current in magnitude correspond to periodic jumps of the transport current penetrating into the ring, and in the intervals between the jumps in both cases the ring is in a superconducting state.

Since there is no quantitative theory of nondissipative current states of a quantum interferometer which has a variety of parallel-connected superconducting point contacts and is shunted by a superconducting inductor, it is only possible to try to explain qualitatively the current states of the ring with a clamping contact. For this we represent the ring with the specified contact in the form of the model diagram shown in Fig. 3. The model diagram assumes that the observed dependence $I_L(I_{H\perp})$ is a result of interference of the unquantized diamagnetic current induced in the ring by the field $H_\perp$ and quantized values of the current $i$ circulating in the interferometer and appearing in it under the action of the secondary field $h$ of the current $I_{H\perp}$. To simplify the analysis of processes occurring in this structure, in this diagram, the clamping point contact is not represented by an interferometer with multiple microcontacts connected in parallel, but by a low-inductive ($L_0 \sim 10^{-13}$ H) interferometer with two Josephson junctions $1$ and $2$ and with different inductances of its branches included in the ring with a large inductance ($L \sim 10^{-6}$ H). The current $I_L$, excited by the magnetic field $H_\perp$ in the ring, creates around niobium wire the secondary magnetic field $h$, acting on the interferometer,

$$h \approx I_L/\pi d. \qquad (1)$$

As known,[10] a magnetic field (in this case the field $h$) induces a periodic modulation of the critical current of an interferometer. The primary reason of the periodic dependence is quantization of the circulating current $i$ in the interferometer, for which the relation $\Phi_0^2/2L_0 > kT$ is fulfilled. In our case of the miniature interferometer, this relation is well fulfilled. In this case a current circulating periodically (with a period in flux equal to half a quantum of magnetic flux in the interferometer part) can be either diamagnetic or paramagnetic. In the first approximation, the depth of quantum modulation of the critical current in the interferometer is equal to a maximum value of the circulating current,[10]

$$i_m = \Phi_0/2L_0. \quad (2)$$

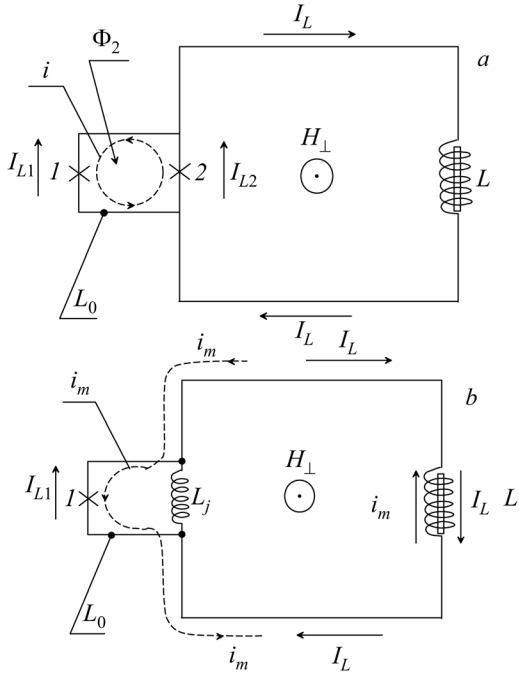

FIG. 3. A model scheme of the ring with a two-contact interferometer in a magnetic field $H_\perp$ and a distribution of the diamagnetic current $I_L$, lower than the critical current of the interferometer, $L$ is the ring inductance, $L_0$ is the interferometer inductance ($L \gg L_0$), $1$ and $2$ are the Josephson contacts, $\Phi_2$ is the magnetic flux through the interferometer, $i$ is the circulating current in the interferometer (a). A model scheme of the ring and a distribution of the currents $I_L$ and $i$ when the sum of the currents ($I_L + i$) achieves a value of the critical current of the contact $2$ (b).

For a maximum value of the circulating current the interferometer critical current is minimal, and maximal for zero value. If the maximum critical current of the interferometer is greater than the depth of its modulation, the diamagnetic current $I_L$, excited by the magnetic field $H_\perp$, at first with increasing the field $H_\perp$ may increase to a maximum value of the critical current of the contact in accordance with the relation

$$I_L = \Phi/L = \mu_0 H_\perp S/L, \quad (3)$$

where $\Phi$ is the magnetic flux of the field $H_\perp$ through the ring area $S$, $\mu_0 = 4\pi \cdot 10^{-7}$ H/m. Then it will decrease according to a certain law down to its minimum value by an amount equal to the depth of modulation $\Delta I_L$. With further increase of the field $H_\perp$ the process becomes periodic. Approximately the same periodic dependence of the critical diamagnetic ring current on the magnetic field can be observed during an experiment (Fig. 2(a)). In this case, from the experimental data and using Eqs. (1) and (2), we can estimate the value of $\Delta I_L$, which appears to be close to its experimental value. A difference between the model dependence and experimental one is in the presence in the experimental dependence of a stepwise decrease of the current $I_L$ in the ring after its reaching of the maximum value and of its strictly linear diamagnetic increase in accordance with Eq. (3) after reaching a minimum value.

Changing the current in the ring with a decrease in the magnetic field $H_\perp$ can be explained by excitation of the ring current in the opposite direction with the achievement of similar periodic critical states of the interferometer in the negative range of the current $I_L$. The decrease of the field out from the values corresponding to first two minima of the critical diamagnetic current clearly demonstrates the quantum nature of freezing the current in portions equal to the depth of modulation of both the diamagnetic and paramagnetic currents in the ring.

Finally, we consider possible causes of appearing a specific stepwise behavior of the periodic dependence of the critical diamagnetic current on the external magnetic field $H_\perp$. To determine these causes some information may be obtained from the dependence of the form $I_L(I_{H\perp})$. In Fig. 2 it is seen that the depth of current modulation, $\Delta I_L$, is approximately three times less than the maximum critical diamagnetic current determined by the critical current of the microinterferometer formed by the clamping contact. Thus, for the model interferometer in Fig. 3(a) with contacts which have similar critical currents $I_c$, it means that the parameter $2I_cL_0/\Phi_0$ is greater than unity. It is known[11] that for these values of the mentioned parameter the dependence of the circulating current on the reduced value of the magnetic flux $\Phi_2/\Phi_0$ ($\Phi_2 \approx \mu_0 hs$, where $s$ is the quantization area of the interferometer) acquires a steplike form in the range of values $n\Phi_2/2\Phi_0$ ($n = 1, 2, 3 \ldots$) when it reaches the critical current of the interferometer. The circulating current abruptly changes the direction from the diamagnetic to the paramagnetic with a value equal to $i_m$. On the other hand, the circulating current $i$ is added in one (for instance, in the second) contact to a part $I_{L2}$ of the ring current $I_L$ (assume that $I_{L2} \approx I_L/2$) and is subtracted form it in the other (first) contact depending on direction of the current $i$, while holding the relation $(I_L/2) + i < I_c$ (Fig. 3(a)). With increasing the field $H_\perp$ and, hence, $i$ there might appear a situation when in one of the contacts of the interferometer (e.g., in the second) the sum $(I_L/2) + i_m$ achieves a value of the critical current $I_c$ of the interferometer contact. At this value of the current the parametric Josephson inductance $L_J$ of the contact increases sharply,[12]

$$L_J = \Phi_0/(2\pi I_c \cos \varphi), \quad (4)$$

where $\varphi$ is the phase difference of wave functions of Copper pairs on the contact depending on the current through the contact. In particular, at $\varphi = \pi/2$ the parametric inductance turns to infinity that physically means depairing of electrons in the contact. Because of strong increase of $L_J$ (to a value greater than the inductance of the ring $L \sim 10^{-6}$ H) in one of the contacts near $I_c$, the paramagnetic circulating current appearing at this moment with an amplitude $i_m$ can be redistributed in such a way that for a short time, comparable to the depairing time equal to $\sim 10^{-12}$,[11] it will close the path, consisting of a part of the interferometer circuit with the first contact, carrying a current less than critical one, and the rest of the ring, which has an inductance of about $10^{-6}$ H (Fig. 3(b)). In this case the paramagnetic current $i_m$ appearing abruptly is subtracted from the diamagnetic ring current that may explain the experimentally observed stepwise decrease in the current $I_L$ (Fig. 2). Because of this process the current of the second contact is reduced. It goes into a subcritical state, as the interferometer also does as a whole. The paramagnetic current, caught in the ring, freezes in it similar to freezing the transport current after the appearance of a current step in the dependence $I_L(I)$ in Ref. 8. Such a freezing does occur in our experiment if to take into account the appearance of the paramagnetic current in the ring after the first and the second jumps of the current $I_L$ (Fig. 2). A transfer of a part of the circulating current from the low-inductance interferometer to the ring with a large inductance $L \sim 10^{-6}$ H occurring within a very short time does not contradict the energy relation, which prohibits the possibility of observing quantum processes in rings with a large inductance $L$, containing the Josephson junction,

$$\Phi_0/2L < kT, \qquad (5)$$

since this relation does not include a time parameter, and applies only to long-term steady states in such superconducting structures.

Returning to the analogy between the dependences $I_L(I)$ and $I_L(I_{H\perp})$ for a ring with the Josephson clamping point contact, it can be said that such a contact is a unique electronic key, either forming in the ring quantized values of the transport current in the form of current steps (in the case of introducing into the contact-interferometer of constant transport current shown in Fig. 1 by a dashed line), or forming quantized values of the paramagnetic current, which are equal to current steps (in the case of excitation of the diamagnetic current in the ring by an external magnetic field).

To develop a quantitative theory of the phenomena described in the work it is necessary to analyze more profoundly physical causes of the strict periodicity and the stepwise change of the dependence $I_L(I_{H\perp})$ with help of both a proposed model of the ring with an interferometer and the experimental data with the ring containing a two-contact interferometer, which has known in advance the dimensions and values of critical currents of contacts.

From a viewpoint of practical use of the results it seems to be promising to investigate possibilities of creating a new type of a quantum magnetometer which is based on a superconducting ring with an interferometer and does not require, in contrast to the known SQUIDs, the supply of energy for its operation.

**CONCLUSION**

Studying a behavior of a superconducting ring with a point Josephson junction in a constant magnetic field allowed to establish a number of analogies with the processes occurring in the same ring but under the action of a constant transport current flowing through it. In particular, we confirmed the conception that a clamping point contact can be represented as a complex quantum microinterferometer the current state of which can vary under the action of the transport current as well as under the action of the diamagnetic current excited in the ring by an external magnetic field. The physical reason for these changes is the interference of either the transport current or the diamagnetic current with the quantum circulating currents of a microinterferometer. Therefore, the adjective "point" for a clamping contact is relative and must be enclosed in quotes.

From the investigations performed it follows that the critical nondissipative current state of a "point" contact in the form of the quantum interferometer with a number of random Josephson microcontacts connected in parallel, which is shunted by a superconducting inductor, may vary strictly periodically as a function of an external magnetic field, regardless of the complexity of its microstructure. At the same time one of the possible causes of the periodic stepwise decrease of the critical current in a superconducting diamagnetic ring with a clamping Josephson "point" contact under the influence of an external magnetic field can be short ($\sim 10^{-12}$ s) and periodic switching of a paramagnetic part of the quantum circulating current in the microinterferometer, formed in the clamping contact, to a branch of the ring with a large inductance ($L \sim 10^{-6}$ H). The latter circumstance can be used as the basis for development of a new type of superconducting quantum magnetometer.